\documentclass[showpacs,pre,aps,superscriptaddress,twocolumn,floatfix]{revtex4}
\usepackage[usenames,dvipsnames]{xcolor}
\usepackage{amsmath}
\usepackage{amssymb,mathrsfs}
\usepackage{graphicx}

\newcommand {\prt} {\partial}
\newcommand {\al} {\alpha}
\newcommand {\vphi} {\varphi}

\begin{document}

\title{On exact solutions of nonlinear acoustic equations}

\author{A. M. Kamchatnov}
\affiliation{Institute of Spectroscopy,
  Russian Academy of Sciences, Troitsk, Moscow, 108840, Russia}
\author{M. V. Pavlov}
\affiliation{Department of Mechanics and Mathematics,
Novosibirsk State University, 2 Pirogova street, Novosibirsk, 630090, Russia}

\date{\today}

\begin{abstract}
Solutions of nonlinear acoustic equations describing propagation of strong
sound pulses with account of curvature of wave fronts in multi-dimensional
geometry are obtained from simple physical considerations. The form of these solutions
suggests {\it ansatz} suitable for finding solutions of much more general
equations of Khokhlov-Zabolotskaya type. General method is illustrated by an
example of nonlinear sound pulse focused in one
transverse direction and defocused in the other direction.
\end{abstract}


\maketitle

\section{Introduction}

Propagation of sound pulses and sound beams in weakly nonlinear media with account of
small curvature of wave fronts is of considerable interest for many applications in
science and industry from geology to medicine (see, e.g. \cite{hb-98}). Mathematical theory of
propagation of nonlinear sound pulses is often based on the study of the so-called
Khokhlov-Zabolotskaya (KZ) equation. If we denote, say, the flow
velocity of the medium at a point $\mathbf{r}$ at a moment $t$ as $u(\mathbf{r},t)$,
then the KZ equation can be written in the form
\begin{equation}\label{eq1}
    (u_t+(c_0+\al u)u_x)_x+\frac12c_0\Delta_\bot u=0,
\end{equation}
where $x$ is the axis of propagation of a sound pulse, $c_0$ is the sound
velocity of linear waves in the medium under consideration, $\Delta_\bot$ is the
Laplace operator in directions normal to the propagation direction, i.e., $\Delta_\bot=\prt_y^2$
in 2D-geometry and $\Delta_\bot=\prt_y^2+\prt_z^2$ in 3D-geometry, and subscripts
denote derivatives with respect to corresponding variables. In fact, this equation
appeared first in the paper \cite{lrt-1948} in the context of the theory of compressible gas flow
past a thin wing but at that time it was not related with the problem of propagation of strong
sound pulses. Later it was derived by A.P.~Sukhorukov in framework of nonlinear
physics studies that were performing in R.V.~Khokhlov group and was published in
\cite{kz-1969}. The name {\it Khokhlov-Zabolotskaya equation} was coined by O.V.~Rudenko
and early history of this equation is described in much detail in his paper \cite{rudenko}.
After the soliton theory arose in 70's, equation (\ref{eq1}) for a particular case
of two spatial coordinated ($x$ and $y$) has drawn much attention as a dispersionless
limit of the famous Kadomtsev-Petviashvili (KP) equation and the name {\it KZ equation}
has become commonly accepted.

Both for applications and mathematical theory of nonlinear waves it is important to have
methods of finding physically reasonable solutions of the KZ equation and of other equations
of the same type. First approximate solutions of this equation were studied already in the
papers \cite{kz-1969} (see also \cite{rs-1977}). Methods related with the complete
integrability of the KZ equation (dispersionless limits of the KP equation) were developed,
for example, in \cite{kodama-88,kodama-89,fk-2004,ms-11} and in these papers some interesting
exact solutions were presented. Another methods based on the contact geometry approach to
nonlinear differential equations were developed in \cite{lych-79,klr-07} and they were applied,
in particular, to the KZ equation and its particular solutions were found. However,
all these methods are mathematically involved and hardly can be used in more general
situations important for applications.

The aim of this paper is to present a simple and direct method of finding exact
solutions of equations of KZ type. At first in section 2 we present derivation
of particular solutions of Eq.~(\ref{eq1})
in two- and three-dimensional cases based on physical reasoning.
These solutions coincide with those found in \cite{ms-11,klr-07} by
much more complicated methods and their form suggests {\it ansatz} for a generalization,
developed in section 3, for finding exact solutions of other equations of this type.
Our approach allows us to construct more general solutions than
those found in the aforementioned papers and in the recent preprint \cite{ss-15}.
In section 4 we illustrate the general theory by particular examples and in the last
section 5 we present conclusions.

\section{Physical derivation of the exact solution of the KZ equation}

Although the KZ equation has a quite general nature, for definiteness we shall imply here
a problem of propagation of a strong sound pulse through a gaseous medium. In the main linear
approximation the pulse propagates with the sound velocity $c_0$. As a variable that describes
the wave motion we choose a local flow velocity $u(x,t)$. If this motion can be described with
a sufficient accuracy as a plane unidirectional wave, then the corresponding wave equation
reduces to
\begin{equation}\label{eq2}
  u_x+c_0u_x=0
\end{equation}
with obvious solution
\begin{equation}\label{eq3}
  u(x,t)=F(x-c_0t),
\end{equation}
where $F(X)$ is an arbitrary function that can be determined from the initial condition.

If we take into account weak nonlinear effects, then the sound velocity becomes dependent on
the amplitude of the wave and the first approximation for the nonlinear correction to velocity
of liquid particles in the wave profile is proportional to the amplitude $u(x,t)$,
\begin{equation}\label{eq4}
  \frac{dx}{dt}=c_0+\alpha u(x,t),
\end{equation}
where the coefficient $\alpha$ can be expressed in terms of the dependence of the pressure $p$
on the specific volume $V=1/\rho$ ($\rho$ being the gas density) as follows (see, e.g.,
\cite{LL-6}):
\begin{equation}\nonumber
  \alpha=\frac{c_0^4}{2V^3}\left(\frac{\partial^2V}{\partial p^2}\right)_s,
\end{equation}
where the derivative is calculated at the condition that the entropy $s$ is constant.
For example, in the case of a polytropic gas we have $\alpha=(\gamma+1)/2$ where
$\gamma=c_p/c_v$ is the heat capacity ratio.

The nonlinear contribution $\alpha u$ to the velocity leads to an additional displacement
of the wave profile points and as a result the solution (\ref{eq3}) should be
modified as
\begin{equation}\label{eq6}
  u=F[x-(c_0+\alpha u)t].
\end{equation}
This equation determines implicitly the dependence of $u=u(x,t)$ on time and space coordinates
for a given function $F(X)$. In other words, the nonlinear correction transforms the linear
equation (\ref{eq2}) into the so-called Hopf equation
\begin{equation}\label{eq7}
    u_t+(c_0+\alpha u)u_x=0
\end{equation}
whose characteristic equation coincides with Eq.~(\ref{eq4}).

For transition to the case of several space dimensions we have to take into account
displacements caused by curvature of the wave pulse front. In a simplest way it can be done
for cylindrical and spherical geometries, and such an additional displacement was calculated
long ago by Landau \cite{landau} and Whitham \cite{whitham} (see also section 102 in \cite{LL-6} and
section 9.1 in \cite{whitham-74}).

In cylindrically symmetric case a linear wave at large enough distance from the origin
(i.e. for $c_0\tau\ll r$, where $\tau$ is the pulse duration) can be represented by the
asymptotic formula
\begin{equation}\label{eq9}
    u\cong\frac1{\sqrt{r}}F(r-c_0t).
\end{equation}
After time $t-t_0=(r-r_0)/c_0$ the pulse propagates due to nonlinear correction to velocity
at the additional distance
\begin{equation}\nonumber
    \delta r=\int_{t_0}^t\alpha udt=\frac{\alpha F}{c_0}\int_{r_0}^r\frac{dr}{\sqrt{r}}=
    \frac{2\alpha F}{c_0}(\sqrt{r}-\sqrt{r_0}).
\end{equation}
Assuming $r\gg r_0$ and taking into account that in the additional term we can make a replacement
$F\sqrt{r}=ur\cong uc_0t$ and also replacing $r$ by $c_0t$ in the coefficient $1/\sqrt{r}$,
we modify (\ref{eq9}) as follows:
\begin{equation}\label{eq11}
    u=\frac1{\sqrt{t}}F[r-(c_0+2\alpha u)t].
\end{equation}
This formula describes evolution of a cylindrically symmetric weakly nonlinear wave.

In a similar way we start from the expression
\begin{equation}\label{eq12}
    u=\frac1r F(r-c_0t)
\end{equation}
for the solution of a spherically symmetric linear wave equation and find the
additional distance of propagation due to the nonlinear correction to the
sound velocity,
\begin{equation}\nonumber
    \delta r=\int_{t_0}^t\alpha u dt=\frac{\alpha F}{c_0}\int_{r_0}^r\frac{dr}r=
    \frac{\alpha F}{c_0}\ln\frac{r}{r_0}\cong \alpha ut\ln\frac{t}{t_0}.
\end{equation}
After replacement with the same accuracy $1/r\to1/(c_0t)$ we arrive at the
expression
\begin{equation}\label{eq14}
    u=\frac1t F\left[r-\left(c_0+\alpha u\ln\frac{t}{t_0}\right)t\right]
\end{equation}
describing evolution of a spherically symmetric weakly nonlinear wave.

We can confirm validity of the expressions (\ref{eq11}) and (\ref{eq14})
by the perturbation theory calculations. As is known (see, e.g., \cite{rs-1977}),
in cylindrically or spherically symmetric geometries evolution of waves
through weakly nonlinear medium is governed by the equation
\begin{equation}\label{eq15}
    u_t+(c_0+\alpha u)u_r+\frac{\kappa}tu=0,
\end{equation}
where $\kappa=1/2$ for cylindrical case and $\kappa=1$ for spherical case. Making
variables replacements
\begin{equation}\label{eq16}
    u=\sqrt{\frac{t_0}t}\, U,\quad T=2\sqrt{t_0t},\quad X=r-c_0t
\end{equation}
in cylindrical case or
\begin{equation}\label{eq17}
    u=\frac{U}t,\quad T=\ln\frac{t}{t_0},\quad X=r-c_0t
\end{equation}
in spherical case, we transform Eq.~(\ref{eq15}) to the Hopf equation
\begin{equation}\label{eq18}
    U_T+\alpha UU_X=0
\end{equation}
with well-known solution
\begin{equation}\label{eq19}
  X-\al UT=F(U)
\end{equation}
that is transformed to (\ref{eq11}) and (\ref{eq14}) after returning to
the initial physical variables.

Now we can turn to the KZ equation (\ref{eq1}). We notice that the solutions
(\ref{eq11}) and (\ref{eq14}) for small values of transverse coordinates $y,\,z$
compared with the propagation distance $x$ must be the solutions of the KZ equation
in the first order of the series expansion with respect to
a small parameter $(y^2+z^2)/x^2\cong(y^2+z^2)/(c_0t)^2$ (term with $z^2$ should
be omitted in the cylindrical case). Thus, we arrive at the solution
\begin{equation}\label{eq20}
    u=\frac1{\sqrt{t}}F\left[x+\frac{y^2}{2c_0t}-(c_0+2\alpha u)t\right]
\end{equation}
of cylindrical KZ equation and
\begin{equation}\label{eq21}
    u=\frac1t F\left[x+\frac{y^2+z^2}{2c_0t}-\left(c_0+
    \alpha u\ln\frac{t}{t_0}\right)t\right]
\end{equation}
of spherical KZ equation. These solutions were found respectively in \cite{ms-11} and
\cite{lych-79,klr-07}
by different more complicated mathematical methods.

On one hand, the presented here derivation shows that solutions of
multi-dimensional equations that take into account small curvature of wave fronts
can be found from corresponding symmetrical solution by their series expansions
with respect to small transverse coordinates. Therefore this method can be easily
generalized on other forms of nonlinearity and number of transverse dimensions.
On the other hand, substitutions (\ref{eq16}) or (\ref{eq17}) suggest that similar
transformations can be done directly in the KZ equation leading to its
exact solutions. We shall develop such a direct method for quite general class
of nonlinear acoustic equations in the next section.

\section{Direct method}

For simplicity of notation, we shall consider here the KZ equation in non-dimensional
form,
\begin{equation}\label{eq22}
  (u_t+u^nu_x)_x+\frac12\Delta_{\bot}u=0,
\end{equation}
for generalized nonlinearity $u^nu_x$ and arbitrary number $N$ of transverse
spatial coordinates $\mathbf{z}=(z_1,\,\ldots,z_N)$,
\begin{equation}\label{eq23}
  \Delta_\bot=\frac{\partial^2}{\partial z_1^2}+\ldots +\frac{\partial^2}{\partial z_N^2}.
\end{equation}
Of course, in standard physical application we have $N=1$ or $N=2$. Besides that,
we shall confine ourselves to the case of integer positive values of $n$.

Solutions (\ref{eq20}) and (\ref{eq21}) as well as substitutions (\ref{eq16})
and (\ref{eq17}) suggest that it might be possible to obtain a wide class
of solutions of the equation (\ref{eq23}) with the use of the following {\it ansatz}
\begin{equation}\label{eq24}
\begin{split}
  &u(t,x,\mathbf{z})=a(T)U(X,T),\\
  &X=x+\vphi(t,\mathbf{z}),\quad T=T(t).
  \end{split}
\end{equation}
In these new variables the equation (\ref{eq22}) takes the form
\begin{equation}\label{eq25}
\begin{split}
  [T_tU_T&+(T_ta_T/a+\tfrac12\Delta_\bot\vphi)U\\
  &+(\vphi_t+\tfrac12(\nabla\vphi)^2+a^nU^n)U_X]_X=0,
  \end{split}
\end{equation}
where subscripts denote derivatives with respect to corresponding variables.
If we assume that
\begin{equation}\nonumber
  T_t=a^n,\quad \vphi_t+\frac12(\nabla\vphi)^2=ka^n,\quad \Delta_\bot\vphi=2a^n(m-a_T/a),
\end{equation}
where $k$ and $m$ are some constants, then Eq.~(\ref{eq25}) transforms to the equation
\begin{equation}\label{eq27}
  (U_T+(U^n+k)U_X+mU)_X=0.
\end{equation}
The term with $k$ can be excluded by means of the additional replacement
\begin{equation}\label{eq28}
  \vphi=\psi(t,\mathbf{z})+kT(t),
\end{equation}
where $\psi$ satisfies the equations
\begin{equation}\label{eq29}
  \psi_t+\frac12(\nabla\psi)^2=0,\quad \Delta_\bot\psi=2a^n(m-a_T/a),
\end{equation}
and evolution of $U(X,T)$ defined in (\ref{eq24}) and evolving according
to Eq.~(\ref{eq27}) is governed now by the
generalized Hopf equation
\begin{equation}\label{eq30}
  (U_T+U^nU_X+mU)_X=0.
\end{equation}

The first equation in (\ref{eq29}) coincides with the Hamilton-Jacobi
equation for a free particle moving 
in $N$-dimensional space. Its particular solution can be easily found by separation
of variables as a sum of ``actions'' corresponding to separate space coordinates.
Most interesting for us solution \cite{ehor} reads
\begin{equation}\label{eq31}
  \psi(t,\mathbf{z})=\frac12\sum_{p=1}^N\frac{z_p^2}{t-t_p},
\end{equation}
where $t_p,$ $p=1,\dots,N$, are integration constants. Then substitution of this
formula in the left-hand side of the second equation (\ref{eq29}) yields
\begin{equation}\nonumber
  a^{n-1}a_T-ma^n=-\frac12\sum_{p=1}^N\frac{1}{t-t_p}.
\end{equation}
Taking into account $a^n=T_t$, $na^{n-1}a_T=(n/a)a_TT_t=n(a_t/a)=(a^n)_t/a^n$ and
introducing $y=a^n=T_t$, we arrive at the Bernoulli equation
\begin{equation}\nonumber
  y'+\left(\frac12\sum_{p=1}^N\frac{1}{t-t_p}\right) y=nmy^2
\end{equation}
that can be solved by a standard method to give
\begin{equation}\label{eq34}
  y=T_t=-Ce^{-nmT(t)}\prod_{p=1}^N|t-t_p|^{-n/2}
\end{equation}
or
\begin{equation}\label{eq35}
  \left(e^{-nmT(t)}\right)_t=nmC\prod_{p=1}^N|t-t_p|^{-n/2},
\end{equation}
where $C$ is an integration constant. If $m=0$ then Eq.~(\ref{eq34})
simplifies to
\begin{equation}\label{eq35b}
  T_t=-C\prod_{p=1}^N|t-t_p|^{-n/2}.
\end{equation}
Thus, we have reduced finding $T(t)$ to
integration of the function in the right-hand side of Eqs.~(\ref{eq35})
or (\ref{eq35b}).
When $T(t)$ is found, $a(t)$ is determined as $a(t)=(T_t)^{1/n}$.

As a result of the above calculations, the variables in (\ref{eq24}) can be
considered as known and it remains to find the solution of the generalized
Hopf equation (\ref{eq30}). Its integration with respect to $X$ gives at once
\begin{equation}\label{eq36}
  U_T+U^nU_X+mU=g(T),
\end{equation}
where $g(T)$ is an arbitrary function to be determined from the initial conditions.
Equation (\ref{eq36}) can be solved by a standard method of characteristics.
Along characteristic curve starting at the point $U_0(X)$ at $T=0$ we have
\begin{equation}\nonumber
  U_T+mU=g(T)
\end{equation}
and this linear differential equation can be easily solved to give
\begin{equation}\label{eq38}
  U=(U_0(X)+G(T))e^{-mT}
\end{equation}
where we denoted
\begin{equation}\label{eq39}
  G(T)=\int_0^Te^{m\tau}g(\tau)d\tau.
\end{equation}
The characteristic curve is determined by the equation
\begin{equation}\nonumber
\begin{split}
  \frac{dX}{dT}=U^n&=(U_0+G(T))^ne^{-nmT}\\
  &=\sum_{q=0}^n{n\choose q}U_0^{n-q}G^q(T)e^{-nmT}
  \end{split}
\end{equation}
and its integration yields
\begin{equation}\label{eq41}
  X=X_0+\sum_{q=0}^n{n\choose q}U_0^{n-q}\int_0^TG^q(\tau)e^{-nm\tau}d\tau.
\end{equation}
Let the initial distribution $U(X)$ at $T=0$ be given by the function $U_0=F^{-1}(X_0)$,
or in implicit form by the function $X_0=F(U_0)$; then exclusion of $X_0$ and $U_0$
from (\ref{eq41}) gives the final result:
\begin{equation}\label{eq42}
\begin{split}
  X&=F(Ue^{mT}-G(T))\\
  &+\sum_{q=0}^n{n\choose q}(Ue^{mT}-G(T))^{n-q}\int_0^TG^q(\tau)e^{-nm\tau}d\tau.
  \end{split}
\end{equation}
This equation determines implicitly $u$ as a function of $x,\,\mathbf{z},\,t$
through variables
\begin{equation}\label{eq42a}
\begin{split}
  &X=x+kT(t)+\frac12\sum_{p=1}^N\frac{z_p^2}{t-t_p},\\
  &T=T(t),\quad u=a(T)U(X,T)
  \end{split}
\end{equation}
in terms of two arbitrary functions $F(U)$ and $G(T)$ which have to be found from the initial
conditions.

\section{Examples}

First of all, let us check that our general formulas reproduce the known
solutions (\ref{eq20}) and (\ref{eq21}).

In case of a cylindrical KZ equation
\begin{equation}\label{kz1}
  (u_t+uu_x)_x+\frac12u_{yy}=0
\end{equation}
written here in standard non-dimensional notation we choose $n=1$, $N=1$, $k=m=0$,
$t_1=0$. Then we have $a(t)=T_t=t^{-1/2}$ and $X=x+y^2/(2t)$. Consequently
$T(t)=2t^{1/2}$ and $U=t^{1/2}u$. Hence Eq.~(\ref{kz1}) has a solution
\begin{equation}\nonumber
  x+\frac{y^2}{2t}-2tu=F^{-1}(t^{1/2}u)
\end{equation}
or
\begin{equation}\label{kz3}
  u=\frac1{\sqrt{t}}F\left(x+\frac{y^2}{2t}-2tu\right)
\end{equation}
which up to notation coincides with Eq.~(\ref{eq20}).

In a similar way in case of spherical KZ equation
\begin{equation}\label{kz4}
  (u_t+uu_x)_x+\frac12(u_{yy}+u_{zz})=0
\end{equation}
we choose $n=1$, $N=2$, $k=m=0$, $t_1=t_2=0$ and obtain $a(t)=T_t=1/t$,
$X=x+(y^2+z^2)/(2t)$; consequently $T(t)=\ln(t/t_0)$, $U=tu$ and
Eq.~(\ref{kz4}) has a solution
\begin{equation}\nonumber
  x+\frac{y^2+z^2}{2t}-t\ln\left(\frac{t}{t_0}\right)u=F^{-1}(tu)
\end{equation}
or
\begin{equation}\label{kz6}
  u=\frac1{t}F\left(x+\frac{y^2+z^2}{2t}-t\ln\left(\frac{t}{t_0}\right)u\right)
\end{equation}
which again up to notation coincides with Eq.~(\ref{eq21}).

\begin{figure}[th]
        \center{\includegraphics[width=8cm]{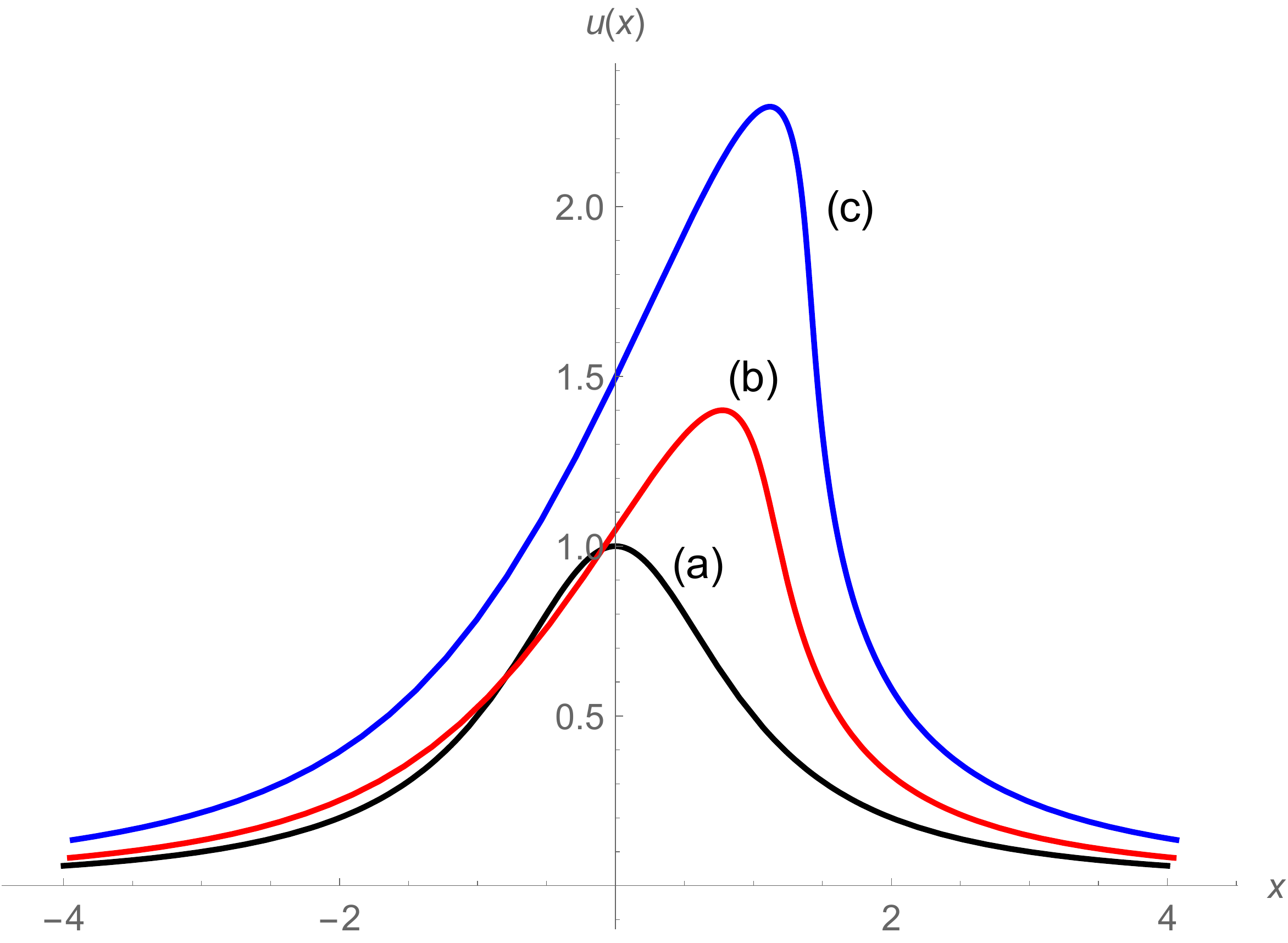}}
        \caption{(Color online) Profiles of the pulse initially given by distribution
        (\ref{eq55}) along $x$ axis for several moment of time: (a) $t=0$;
        (b) $t=0.7$; (c) $t=0.9$. Amplitude tends to infinity as $t\to 1$.}
        \label{fig}
    \end{figure}
Now we notice that
the constants $t_p$ in (\ref{eq31}) and other formulas can have either sign
what means that we can consider sound pulses focused in some transverse directions
and defocused in the other directions. Exact solutions of the KZ equation
for such situations, to the best of our knowledge, have not been considered
earlier and we shall apply here our approach to a problem of this kind.

Thus, we wish to find the solution of the KZ equation (\ref{kz4})
modeling propagation of a nonlinear sound pulse that is defocused in
$y$ direction and focused in $z$ direction.
To this end, we take in the above formulas $n=1$, $N=2$, $m=0$, $k=0$,
$t_1<0$, $t_2>0$, $g(T)=0$, so that
\begin{equation}\label{eq44}
  a(t)=\frac{dT}{dt}=\frac1{\sqrt{(t+|t_1|)(t_2-t)}}
\end{equation}
and, consequently,
\begin{equation}\label{eq45}
  T(t)=2\left(\arctan\sqrt{\frac{t+|t_1|}{t_2-t}}-\arctan\sqrt{\frac{|t_1|}{t_2}}\right),
\end{equation}
where the integration constant is chosen in such a way that $T(0)=0$.
We assume here that $0\leq t<t_2$. The self-similar variable has now
the form
\begin{equation}\label{eq46}
  X=x+\frac12\left(\frac{y^2}{t+|t_1|}-\frac{z^2}{t_2-t}\right).
\end{equation}
The variable $u(x,y,z,t)$ is expressed in terms of $U(X,T)$ as
\begin{equation}\label{eq47}
  u(x,y,z,t)=a(T)U(X,T)=\frac{U(X,T)}{\sqrt{(t+|t_1|)(t_2-t)}},
\end{equation}
where $U$ obeys the Hopf equation
\begin{equation}\label{eq48}
  U_T+UU_X=0.
\end{equation}
We assume that at $t=0$ the distribution of $u_0(x,y,z)$ depends on a single
self-similar variable
\begin{equation}\label{eq49}
  u_0(x,y,z)=F^{-1}\left(x+\frac12\left(\frac{y^2}{|t_1|}-\frac{z^2}{t_2}\right)\right).
\end{equation}
Then the solution of Eq.~(\ref{eq48}) can be written as $X-UT=F(U)$ or,
returning to the original variables,
\begin{equation}\label{eq50}
\begin{split}
  x&+\frac12\left(\frac{y^2}{t+|t_1|}-\frac{z^2}{t_2-t}\right) -2{\sqrt{(t+|t_1|)(t_2-t)}}\\
  &\times\left(\arctan\sqrt{\frac{t+|t_1|}{t_2-t}}-\arctan\sqrt{\frac{|t_1|}{t_2}}\right) u\\
  &=F(\sqrt{(t+|t_1|)(t_2-t)/(|t_1|t_2)}\,u).
  \end{split}
\end{equation}
This formula determines implicitly $u$ as a function of space coordinates
at any moment of time $t$ in the interval $0\leq t<t_2$. It is worth noticing that this restriction makes
it impossible to take the limit $t_1=t_2=0$ and to reproduce the solution (\ref{kz6}).

It is convenient to represent this solution in a parametric form. To this end
we notice that the initial $u_0$ given in the form (\ref{eq49}) is a function
of a single variable
\begin{equation}\label{eq51}
  X_0=x+\frac12\left(\frac{y^2}{|t_1|}-\frac{z^2}{t_2}\right).
\end{equation}
The corresponding value $U=u_0/a(0)=\sqrt{|t_1|t_2}u_0$ is constant along
characteristic $X=X_0+UT$, that is
\begin{equation}\label{eq52}
\begin{split}
  x&+\frac12\left(\frac{y^2}{t+|t_1|}-\frac{z^2}{t_2-t}\right)=
  X_0+2\sqrt{|t_1|t_2}\\
  &\times\left(\arctan\sqrt{\frac{t+|t_1|}{t_2-t}}-\arctan\sqrt{\frac{|t_1|}{t_2}}\right)u_0(X_0).
  \end{split}
\end{equation}
On this surface in 3D-space for fixed $t$ the dependent variable $u$ takes the value
\begin{equation}\label{eq54}
  u=\sqrt{\frac{|t_1|t_2}{(t+|t_1|)(t_2-t)}}\,u_0(X_0).
\end{equation}
Thus the above solution is parameterized by a parameter $X_0$ on which depend both $u$ and
the self-similar variable $X$.
Exclusion of $X_0=F(u\sqrt{(t+|t_1|)(t_2-t)/(|t_1|t_2)}$ and $u_0(X_0)$ from (\ref{eq51})
and (\ref{eq54}) reproduces Eq.~(\ref{eq50}). As we see, in the limit $t\to t_2$ the
amplitude goes to infinity as $u\propto 1/\sqrt{t_2-t}$ due to focusing.

We illustrate this evolution for the case of the initial distribution
\begin{equation}\label{eq55}
  u_0(x,y,z)=\frac1{1+\left(x+\frac12\left({y^2}-{z^2}\right)\right)^2}.
\end{equation}
The profile of the pulse along the $x$-axis for $y=0$, $z=0$ is shown in Fig.~1 for several
moments of time. On the contrary to evolution of defocused pulse, now its amplitude
increases at the axis and the profile steepens due by virtue of nonlinear effects.

\section{Conclusion}

We have demonstrated in this paper that exact nonlinear solutions of the KZ type equations can be
obtained in framework of a simple enough {\it ansatz} leading to partial separation of
variables: evolution in transverse direction reduces effectively to free propagation of
rays that is governed by simple Hamilton-Jacobi equation whereas nonlinear effects are
described by ordinary differential equation that can be integrated in a closed form.
It is important that our method is effective not only in case of completely integrable
situations, when the KZ equation represents a dispersionless limit of the KP equation,
but also in multi-dimensional geometries and generalized nonlinearities. This allows one
to obtain exact solutions in many realistic situations.

From physical point of view, this method combines nonlinear effects with linear
diffraction for wave fronts described by multi-dimensional paraboloid surfaces with
arbitrary signs and values of curvature radii. In practically most important 3D case,
this yields the exact solutions for focused/defocused pulses with engineered phase fronts.
One may hope that such solutions can find many applications in science and technology.

\subsection*{Acknowledgements}

AMK is grateful to S.V.~Manakov for inspiring discussion of the paper \cite{ms-11}.
Work of AMK was partially supported by RFBR (grant No. 16-01-00398).
MVP thanks V.V. Lychagin, who drew his
attention to the results cited in \cite{lych-79,klr-07}, and I. Yehorchenko,
who drew his attention to the results published in \cite{ehor}.
Work of MVP was partially supported by the Russian Science Foundation (grant No.
15-11-20013).

\end{document}